# Mechanical characterisation of the temporomandibular joint disc through local compression and traction

L.K. Tappert[1a], A. Baldit[1b], R. Do Nascimento[1,2], P. Lipinski[1] and R. Rahouadj[1]

[1]LEM3 - UMR CNRS 7239, University of lorraine - ENIM - ENSEM, 1 route d'Ars Laquenexy, BP 65820, 57078 Metz Cedex 3, France, [2]São Carlos Institute of Physics, University of São Paulo, USP, PO Box 369, 1356-6590, São Carlos, SP, Brazil

[a]lara-kristin.tappert@univ-lorraine.fr presenting author and [b]adrien.baldit@univ-lorraine.fr corresponding author

**Abstract.** Temporomandibular joint plays a major role in everyday life allowing speaking and eating. This work aims to describe a method that combine several mechanical tests highlighting soft tissue properties, both local and global, through local compression and traction. It provides "first order" evaluation of mechanical properties and allow setting optimisation loop with more complex models.

## Introduction

Temporomandibular joint plays a major role in everyday life allowing speaking and eating. Many events can affect its behaviour while lifespan therefore a better understanding of the mechanical behaviour of its components, as well as their interactions, is required [1,2]. This work is focused on the soft element of this joint: the disc. It is a biphasic material with up to 66 − 80% of water [3] and mainly made of elastin and collagens type I and II. This composite like structure confers to the tissue an anisotropic, non-linear and viscous behaviour that has to be mechanically characterised to allow simulating the whole articulation comportment. Although, its complex behaviour leads to a large range of mechanical properties [2-6] that mismatch with the aim of predicting its evolution through planned events like surgery. This work aims to describe a method that combine several mechanical tests highlighting soft tissue properties, both local and global, through local compression and traction.

## Material and methods

Porcine temporomandibular discs (Figure 1a) were extracted from slaughterhouse remaining parts and directly frozen within physiological solution until experimentation. The tests performed within physiological solution were sequenced as: i) first local spherical compressions [2], ii) tensile sample extraction and iii) tensile test through fibres directions. Each test aims to characterise a specific property of the tissue. Actually, local compressions in direction perpendicular to fibres reflect the composite like matrix behaviour. Then, parallelepiped samples were extracted from the central disc structure along fibres directions with 9.9x2.1x4.1 mm³ final dimensions measured thanks to a calliper. Eventually, cyclic tensile tests were performed thanks to a Zwick *Zwicki 0.5* tensile machine equipped with a 100 N load cell and driven by displacement during 3 cycles reaching an imposed strain near 10% at a strain rate of $\dot{\varepsilon} \sim 8\times10^{-3}\ s^{-1}$ similar to [2]. This final test aims to characterise the fibre effect on tissue behaviour. The Young's modulus has been computed through *Python* curve fitting on the last measurement points' quarter of each loading phases to observe viscosity effect on this elastic property.

## Results

The tensile sample's extraction highlighted the pre-constrained condition of the disc structure as shown in Figure 1a with an opening angle θ increasing progressively while tissue relaxed and reaching 35°.

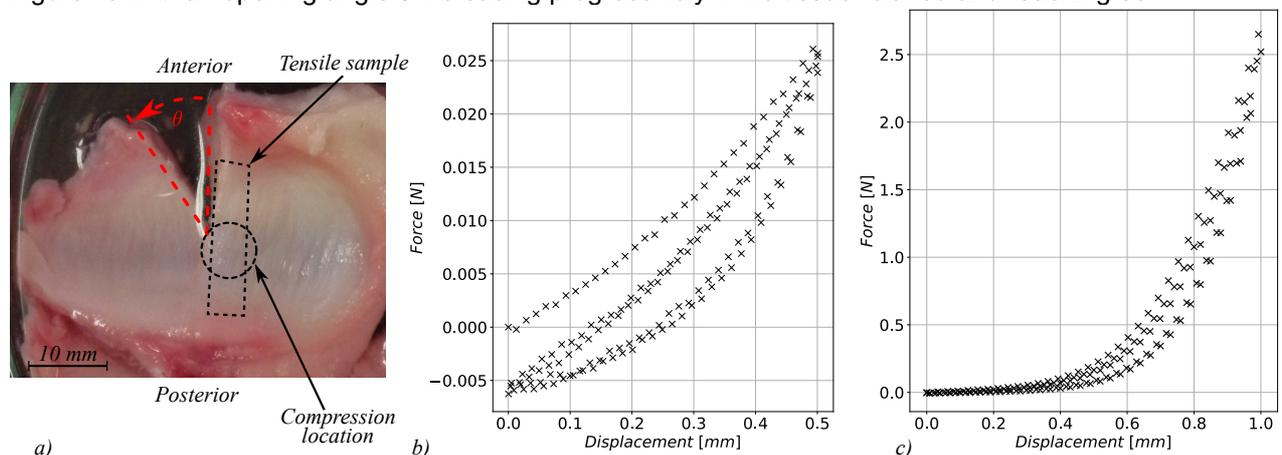

Figure 1: a) Porcine temporomandibular disc with an antero-posterior cut related to the tensile test preparation, b) spherical compression and c) tensile cyclic force-displacement curves.

Both tensile and spherical compression curves had the same characteristics: nonlinear with an accommodation and hysteresis along the cycles as presented in Figures 1b and 1c. Being located at the disc centre, those results are related to the same location with a spherical radius of 3 mm for the compression presented in Figure 1b. In Figure 2a, the spherical compression test presents a Young's modulus of 24.31 kPa and a Poisson's ratio of 0.5. On the other hand, the tensile test allows characterizing the Young's modulus of 9.09 MPa at final loading phase.

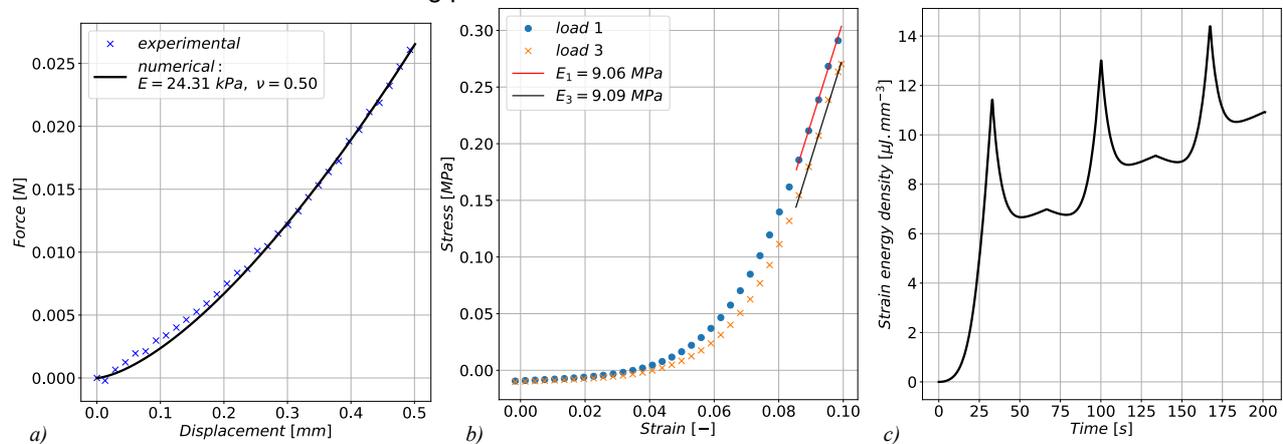

Figure 2: a) spherical compression and b) tensile characterisations and c) strain energy density as function of time during cyclic tensile test.

The tensile test strain energy density presents a non-linear evolution with an accommodation following the stress-strain curve shape and decreasing values for each loading cycle applied.

**Discussion**

Biological tissue growing is done under mechanical constrains leading to an optimised musculoskeletal structure. Tissue residual stresses have been released and observed (Figure 1a) during sample extraction but are not taken into account in the mechanical analysis. In Figures 1b and 1c, the complex mechanical behaviour of the disc tissue is presented under compression and traction loads. The mechanical properties obtained from local compressions are consistent with the literature [2,5]. It characterises the tissue through a quasi-incompressible behaviour reflected by a Poisson's ratio close to 0.5 and a low Young's modulus value related to the matrix made of collagen type II and elastin. The Young's modulus obtained through tensile cyclic test is as well comparable to the literature [4,6] with two orders of magnitude higher value than those obtained from compression tests. This strong variation of elastic properties highlights the anisotropy of the material due to unidirectional orientation of collagen type I fibres in central area of the disc. The strain energy density evolution points out the dissipative behaviour of the tissue.

**Conclusion**

Eventually, the complex behaviour of disc structure requires combined methods to be characterised. Simple methods like Hertz contact theory [2] and linear characterisation provide "first order" evaluation of mechanical properties and allow setting optimisation loop with more complex models [1]. This procedure can be improved using more experimental data through digital image correlation during tensile test [7] and hyper elastic model to characterise more accurate tissue's mechanical properties. The evaluation of the residual stresses through image analyses (as described in [8]) is currently under investigation.

**References**

[1] V. Creuillot, C. Dreistadt, K.J. Kaliński, and P. Lipinski: *Mechatronic Design Towards Investigation of the Temporo-Mandibular Joint Behaviour in Mechatronics: Ideas, Challenges, Solutions and Applications*, edited by J. Awrejcewicz, K.J. Kaliński, R. Szewczyk, and M. Kaliczyńska Publications/Springer International Publishing, 2016 p. 15-32

[2] L.K. Tappert, A. Baldit, R. Rahouadj and P. Lipinski: *Local elastic properties characterization of the temporo-mandibular joint disc through macro-indentation*, Computer Methods in Biomechanics and Biomedical Engineering, Vol. 20 (2017) p. 201-202

[3] M. S. Detamore and K. A. Athanasiou: *Motivation, Characterization, and Strategy for Tissue Engineering the Temporomandibular Joint Disc*, Tissue Engineering, Vol. 9 (2004) p. 1065-1087

[4] M.S. Detamore and K.A. Athanasiou: *Tensile Properties of the Porcine Temporomandibular Joint Disc*, ASME. Journal of Biomechanical Engineering, Vol. 125 (2003) p. 558-565

[5] K.W. Kim, M.E. Wong, J.F. Helfrick, J.B. Thomas and K.A. Athanasiou: *Biomechanical Tissue Characterization of the Superior Joint Space of the Porcine Temporobibular Joint*, Annals of Biomedical Engineering, Vol. 31 (2003) p. 924-930

[6] E. Tanaka, D.P. Rodrigo, M. Tanaka, A. Kawaguchi, T. Shibazaki and K. Tanne: *Stress analysis in the TMJ during jaw opening by use of a three-dimensional finite element model based on magnetic resonance images*, International Journal of Oral and Maxillofacial Surgery, Vol. 30 (2001) p. 421-430

[7] A. Baldit, D. Ambard, F. Cherblanc and P. Royer: *Experimental analysis of the transverse mechanical behaviour of annulus fibrosus tissue*, Biomechanics and Modeling in Mechanobiology, Vol. 13 (2014) p. 643-652

[8] S. Le Floch, D. Ambard, A. Baldit, P. Kouyoumdjian, P. Canadas and F. Cherblanc: *Residual strain and stress in pig intervertebral disc*, Procedings of European Society of Biomechanics 2015.